\documentclass[twocolumn,epjc3]{svjour3}

\DeclareSymbolFont{matha}{OML}{txmi}{m}{it}
\DeclareMathSymbol{\varv}{\mathord}{matha}{118}
\usepackage{amssymb}
\usepackage{xcolor}
\usepackage{soul,ulem}
\usepackage{physics}
\usepackage{upgreek,textgreek}

\smartqed  
\RequirePackage{graphicx}

\RequirePackage{latexsym}
\RequirePackage[numbers,sort&compress]{natbib}
\RequirePackage[colorlinks,citecolor=blue,urlcolor=blue,linkcolor=blue]{hyperref}

\usepackage{commath} 
	\usepackage{braket}  
	\usepackage{lipsum} 
	\usepackage{import}
	\usepackage{color}
\usepackage{bm}
\usepackage{dcolumn}
\usepackage[utf8]{inputenc}
\usepackage[absolute,overlay]{textpos}
\usepackage{hyperref}
\usepackage{graphicx}
\usepackage{dcolumn}
\usepackage{xcolor}
\usepackage[numbers,sort&compress]{natbib}
\usepackage[all]{nowidow}
\usepackage{booktabs,multirow}

\let\oldAA\AA
\renewcommand{\AA}{\text{\normalfont\oldAA}}

\newcommand{\blu}[1]{{\color{black}{#1}}}

\journalname{Eur. Phys. J. Plus}
\begin{document}

\title{Many-Body Dissipative Particle Dynamics Simulations of Lipid Bilayers with the MDPD-MARTINI Force-Field
}



\author{Natalia Kramarz\thanksref{e1,addr1,eq}\and Lu\'is H. Carnevale\thanksref{e2,addr1,eq}
        \and
        Panagiotis E. Theodorakis\thanksref{e3,addr1} 
}

\thankstext{e1}{e-mail: kramarz@ifpan.edu.pl}
\thankstext{e2}{e-mail: carnevale@ifpan.edu.pl}
\thankstext{e3}{e-mail: panos@ifpan.edu.pl}
\thankstext{eq}{Equal contribution}


\institute{Institute of Physics, Polish Academy of Sciences, Al. Lotnik\'ow 32/46, 02-668 Warsaw, Poland \label{addr1}
}

\date{Received: date / Accepted: date}

\maketitle

\begin{abstract}
Many-body dissipative particle dynamics \\
(MDPD) offers a significant speed-up in the simulation
of various systems, including soft matter, 
in comparison with molecular dynamics (MD) simulations based on Lennard-Jones interactions, 
which is crucial for describing phenomena
characterized by large time and length scales. Moreover,
it has recently been shown that the MARTINI force-field
coarse-graining approach is applicable in MDPD,
thus rendering feasible the simulation of
complex molecules as in MD-MARTINI for ever
larger systems and longer physical times.
Here, simulations of various lipid membranes
were performed by using the MDPD-MARTINI coarse-grained (CG) force-field (FF), 
relevant properties were calculated, and comparison with 
standard MD-MARTINI CG simulations and experimental data
was made. 
Thus, insights into structural properties of these bilayer systems and 
further evidence regarding the transferability of the 
MDPD-MARTINI CG models are provided.
In this regard, this is a natural next step 
in the development of the general-purpose MDPD-MARTINI-FF, 
which generally provides significant speed-ups 
in both computational and physical simulated times,
in comparison with standard CG MD simulations. 
\end{abstract}

\section{Introduction}
Molecular dynamics (MD) \cite{Rapaport_2004} is a common method used for the
simulation of a whole range of materials in various areas, such as soft matter,
fluid dynamics, and solid-state physics. A key element for reliable MD simulations
is the ability of describing the interactions between particles (force field) in such a way that
various experimental properties of interest of a system are well-reproduced by the simulation.
To this end, the non-bonded interactions between particles play a key role in
reproducing these properties and much of the research for creating the various force-fields (FFs)
is mainly focused on adequately describing those non-bonded interactions for a given
cutoff at a specific temperature (the development of FFs valid over a wide range of
temperatures is an active field of research), with further
refinements coming from the topology FF parameters, which often can also
be crucial.
When interactions are considered between all atoms of the system, that is in the
case of all-atom FFs, simulations are generally computationally expensive, 
albeit often necessary, for example, when a finer time- and length-scale resolution
is required, what has led to the development of coarse-grained (CG) FFs \cite{Noid2013,Kmiecik2016,Jin2022,Noid2023}. 
In this case, 
interactions between \blu{groups} of atoms are \blu{modelled} instead of individual atoms, thus the number
of required force (interactions) calculations during the integration of
the equations of motion in MD simulations is significantly reduced,
which, in turn, results in faster simulations and significant savings
in energy resources consumed to run the simulations and cool down the hardware.
Moreover, CG models offer the possibility of using a larger time step in the simulation
and faster dynamics, which provides an additional speed-up of the simulations.

MARTINI is one of the most popular CG FFs for MD simulation, since it is a general-purpose
FF that can be applied in the simulation of a wide spectrum of 
components \cite{Marrink2004,Marrink2007,Marrink2013,Alessandri2021},
such as proteins \cite{Monticelli2008,Periole2009,deJong2013,Poma2017},
polymers \cite{Lee2009}, carbohydrates \cite{Lopez2009},
glycolipids \cite{Lopez2013}, glycans \cite{Chakraborty2021},
DNA \cite{Uusitalo2015}, RNA \cite{Uusitalo2017}, water \cite{Yesylevskyy2010}
and various solvents \cite{Vainikka2021}. In addition, various extensions include
simulations for specific pH \cite{Grunewald2020} and more recently chemical reactions
(reactive MARTINI) \cite{Sami2023}. While the MARTINI 3.0 CG FF \cite{Souza2021} 
has recently been released, the development of any FF is a continuous effort that 
aims at improving the FF parameters through refinement based on a wider range
of methodologies and validation with experimental data
on an ever growing number of possible systems. 
In the case of CG FFs, validation can often take place by comparison
to results obtained by well-established all-atom FFs or CG models, which 
have already been parametrized according to experimental data.

Regarding the MD-MARTINI FF, its popularity is mainly due to the fact that it allows for
the simulation of a wide spectrum of different systems by only using a finite number
of CG bead types and a top-down approach for non-bonded interactions. 
Here, the concept of a 'LEGO' approach of combining these beads to
build more complex molecules while accounting for chemical specificity is used. 
In addition, MARTINI is based on the Lennard-Jones potential 
with the smallest possible cutoff including attractions, which, in principle, 
renders this model computationally efficient, while various features
implemented in GROMACS software (e.g. virtual atoms) \cite{Berendsen1995},
namely the native platform for MARTINI,
allow for further speed-up of the simulations. Currently, the MD MARTINI method
enjoys a large community, which applies and further tests the models in a range
of diverse systems, thus practically participating in the
validation effort of this general-purpose FF.

Despite the wide use of the MD-MARTINI model, 
MD remains a computationally expensive method even for CG models.
However,
it has recently been shown that the MARTINI approach can be applied in the case of
many-body dissipative particle dynamics (MDPD) \cite{Carnevale2024_MDPD_MARTINI}, a method
that offers a computational
speed-up of about an order of magnitude in comparison with the MD method due to the short-range interactions, 
as well as an additional speed-up from the use of
soft interactions, while at the same time
it accounts for both repulsion and attraction between particles
\cite{Pagonabarraga2001,warren2003,zhao2017,zhao2021-review,zhao2021,Han2021}.
\blu{In particular, it has been shown that the MDPD simulations are 4--7 times faster than the
MD simulations for systems of DPPC lipids and water \cite{Carnevale2024_MDPD_MARTINI}. Other comparisons in the literature
have shown that MDPD models are four times faster than MD MARTINI models \cite{Zhu2021}, while in simpler systems
(\textit{e.g.} only water) MDPD models perform even faster \cite{Ghoufi2011}. }
Moreover, there is
no limitation in combining MDPD and MD models in the
same simulation. However, such a scenario is in general
undesirable, since it will naturally lead to a higher 
computational cost for the simulation.
Here, we build on previous work that dealt with the study of DPPC bilayers and further apply the
MDPD-MARTINI-FF in the investigation of POPC and DOPC lipid
bilayers and compare the obtained results with those
of standard MD-MARTINI simulations. Such a comparison 
between the MD- and MDPD-MARTINI FFs is realized on
the basis of a number of related properties for bilayer
systems. Moreover, a new interaction level is provided 
to account for the C3 MARTINI
bead-type in DOPC and POPC lipids. 
The MDPD-MARTINI models \blu{provide} equivalent results
to the MD MARTINI models, while they offer at the same
time a significant speed-up in the simulations, 
namely several minutes of MDPD simulations
instead of several hours of MD simulations, while
interactions are transferable for these systems
as in the case of the MD-MARTINI-FF.

In the following, we present the MDPD- and MD-MARTINI models for the various
lipid systems and details on the simulation methods. Then, we present
the analysis and discussion of various
properties. 
In the last section, we draw our conclusions and discuss next
steps in the development of the general-purpose
MDPD-MARTINI FF.

\section{Simulation models and methods}\label{methods}

\begin{figure}[t]
    \centering
    \includegraphics[width=\linewidth]{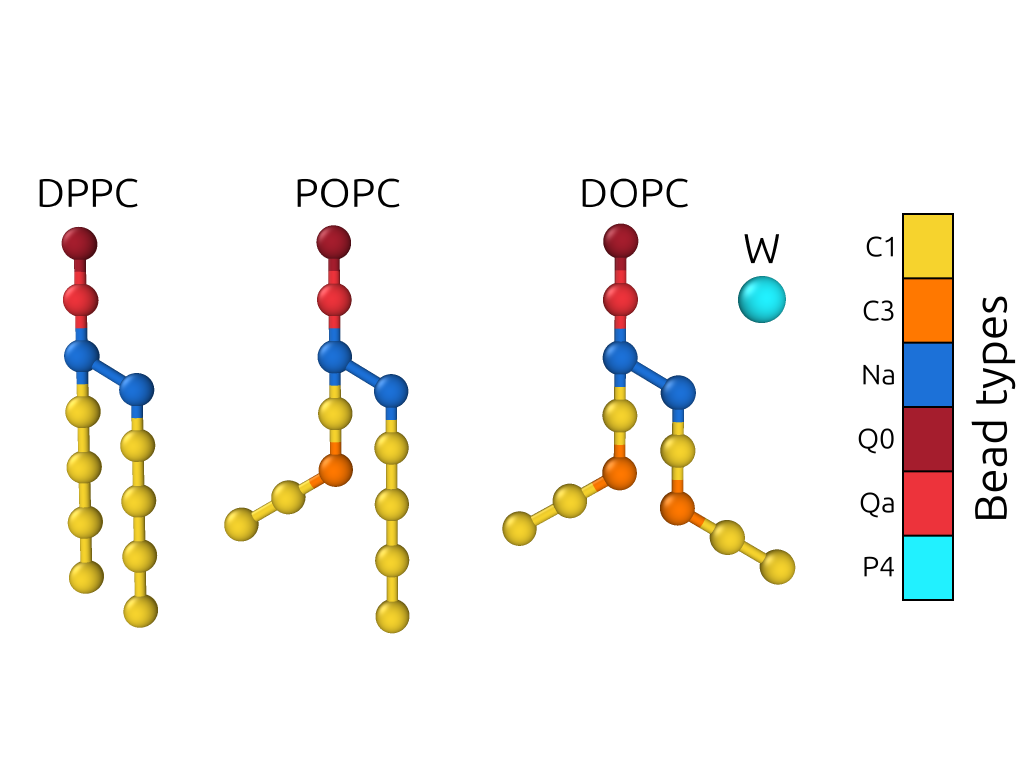}
    \caption{MD and MDPD models of DPPC, DOPC, and POPC lipids. The type of beads are indicated by the different color.
    In particular, $Q_0$ represents the 
    choline headgroup, $Q_a$ the phosphate headgroup, 
    two $N_a$ beads correspond to the glycerol ester moiety, 
    while the two alkane chains consist of $C_1$ and $C_3$ 
    (for POPC and DOPC) beads. The POPC has a double bond
    at carbon atom positions 9 and 10, while the DOPC has a double 
    bond at each chain, as illustrated. 
    The interactions between the beads
    are given in Table~\ref{interaction_matrix}.}
    \label{Lipids_models}
\end{figure}

\subsection{Molecular Dynamics}
The MD model follows that of previous work for 
DPPC bilayers \cite{Carnevale2024_MDPD_MARTINI}, which has been
a benchmark system in the case of the MD-MARTINI-FF \cite{lipids_tutorial}
and therefore further details will be here omitted.
Initial files for LAMMPS software \cite{Plimpton1995,Thompson2022} 
for the DPPC, DOPC, and POPC bilayers based on the MD-MARTINI model
can be found as moltemplate examples \cite{lipids_tutorial_moltemplate}.
The DPPC model consists of different bead types, that is 
$Q_0$ for the choline headgroup, $Q_a$ for
the phosphate headgroup, two $N_a$ beads for 
the glycerol ester moiety, and two chains, 
each comprised of four $C_1$ beads, to represent the alkane tails. 
For DOPC and POPC lipids, the model differs from that of DPPC 
lipids. This distinction arises due to the presence of double bonds in 
the alkane chains, with one double bond in POPC and two in DOPC. 
These double bonds require a different angle in the chain 
structure and an additional bead type, C3, to accurately represent 
their molecular geometry.
The lipid models are illustrated in \textbf{Figure \ref{Lipids_models}}, 
with the names of the bead types being the same for both the MD and
MDPD MARTINI models following the MARTINI FF naming
convention. Hence, this implies that
both models use the same mapping
of CG beads to chemical groups of atoms.
Finally, water is represented by the P4 bead type.

\begin{figure*}[t]
    \centering
    \includegraphics[width=0.7\linewidth]{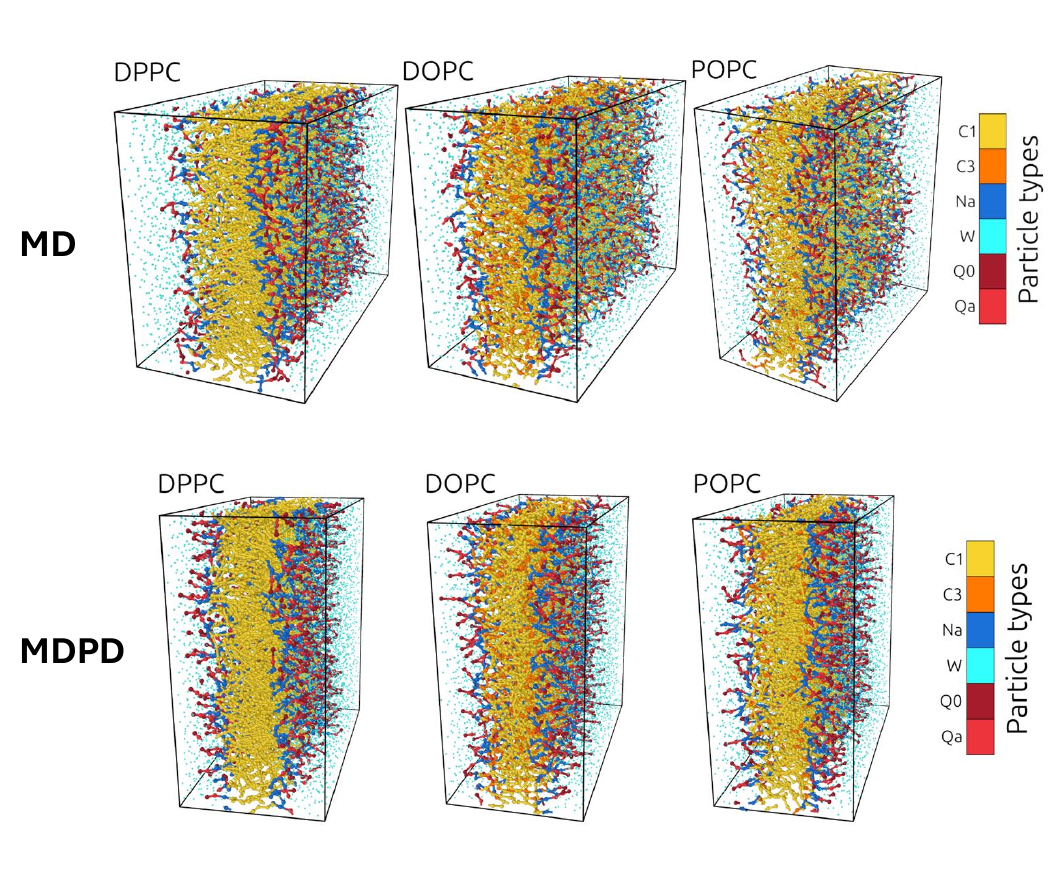}
    \caption{Typical snapshots of DPPC, DOPC, and POPC bilayers of 512 lipids in water. Upper panel corresponds to results
    obtained by MD-MARTINI FF simulations, while the lower panel 
    refers to simulated structures obtained via the MDPD-MARTINI
    force-field. }
    \label{MD_MDPD_Bilayers}
\end{figure*}

Typical simulations for forming the lipid bilayers follow 
the same protocol as in previous work \cite{Carnevale2024_MDPD_MARTINI}. 
However, here, the aim
is not to show the computational speed-up of the method during
the bilayer formation,
since this is well-established for MDPD simulations
and becomes even
apparent during the calculation of the
system properties
after the system has reached equilibrium \cite{Carnevale2024_MDPD_MARTINI},
but rather to assess the quality 
performance of the MDPD-MARTINI model in
the simulation of  DOPC and POPC bilayers with an 
additional C3 bead-type and investigate key properties of
these bilayers.
Typically, the simulated systems consist of 512 lipids, which 
are placed in a simulation box of dimensions 
$L_x=L_y=130$ and $L_z = 100$, maintaining a 16:1 proportion
of water beads.

Preparing the bilayers involves minimizing the system's energy, 
conducting a simulation in the NPT ensemble employing 
the Nos\'e--Hoover thermostat
at 300 K and 1 bar to obtain the equilibrium box volume. 
Then, simulations are carried out at an elevated temperature 
(typically around 450 K) to facilitate the removal of any 
undesired structures that may have formed, and the final
bilayer formation. Further refinement of the system
can take place by realizing NPT simulations employing a 
zero surface tension condition while gradually lowering the 
temperature to 300~K. Equilibrium properties are calculated
by NPT runs at this temperature with typical snapshots for each
system presented in \textbf{Figure \ref{MD_MDPD_Bilayers}}.

\subsection{Many-body dissipative particle dynamics}

The MDPD method has been applied
for fluids with different properties \cite{Espanol1995,warren2003,zhao2017,zhao2021-review,zhao2021,Han2021,Vanya2018,Carnevale2023,Ng2025},
as well as a range of soft matter multi-phase and multi-component
systems, such as systems with surfactant molecules \cite{Hendrikse2023,Carnevale2024}.
The most common use of MDPD simulation is usually carried out
by solving the Langevin equation of motion for each particle
(\textbf{Equation~\ref{eq1}}) as is also done in the case of MD simulations \cite{Theodorakis2011}. 
However, in the case of MDPD, forces are directly defined, instead of
being provided by a potential form.
The positions and velocities of each particle evolve according to 
Equation~\ref{eq1}
\begin{eqnarray}
	m\frac{d\bm{v}_i}{dt} = \sum_{j\neq i} \bm{F}_{ij}^C + \bm{F}_{ij}^R + \bm{F}_{ij}^D,
	\label{eq1}
\end{eqnarray}
which considers the interactions of a
particular particle $i$ with all surrounding particles within a cutoff
distance, $\bm{F}^C$, and the presence of dissipative, $\bm{F}^D$,
and random forces, $\bm{F}^R$, to act as a thermostat.
The mass, $m$, is the same for all particles and set to unity.

The total conservative force on particle $i$, namely,
$\bm{F}^C$, which is the main difference between
dissipative particle dynamics (DPD) and 
MDPD \cite{Espanol2017,Yoshimoto2013,Li2014,Lavagnini2021,Trofimov2005}
is mathematically expressed by the following relation 
\begin{eqnarray}
	\bm{F}^C_{ij} =  A\omega^C(r_{ij})\bm{e}_{ij} + 
	B \left(\bar{\rho_i} + \bar{\rho_j} \right) \omega^d(r_{ij})\bm{e}_{ij}
	\label{eq2}
\end{eqnarray}
and includes attractive ($A<0$) and repulsive ($B>0$) interactions.
$r_{ij}$ is the distance between particles, 
$\bm{e}_{ij}$ the unit vector connecting particles $i$ and $j$,
while $\omega^C(r_{ij})$ and $\omega^d(r_{ij})$ are linear
weight functions given as
\begin{eqnarray}
	\omega^{C}(r_{ij}) = 
	\begin{cases}
		&1 - \frac{r_{ij}}{r_{c}}, \ \ r_{ij} \leq r_{c} \\
		& 0,  \  \ r_{ij} > r_{c}.
	\end{cases} 
	\label{eq3}
\end{eqnarray}
$r_c$ is a cutoff distance for the interactions, 
which is usually set to unity. Moreover, $\omega^d(r_{ij})$ has the same form,
but its cutoff distance is $r_d=0.75$.
The repulsive term in the conservative force depends on the local
densities, which are expressed as follows:
\begin{eqnarray}
	\bar{\rho_i} = 
	\sum_{0<r_{ij}\le r_d}
	\frac{15}{2\pi r_d^3} \left( 1 - \frac{r_{ij}}{r_d}\right)^2 .
	\label{eq4}
\end{eqnarray}
The random and dissipative forces are
\begin{eqnarray}
	\bm{F}^D_{ij} = -\gamma \omega^D(r_{ij}) (\bm{e}_{ij} \cdot  \bm{v}_{ij})\bm{e}_{ij} ,
	\label{eq5}
\end{eqnarray}
\begin{eqnarray}
	\bm{F}^R_{ij} = \xi \omega^R(r_{ij}) \theta_{ij} \Delta t^{-1/2} \bm{e}_{ij} ,
	\label{eq26}
\end{eqnarray}
respectively, with $\gamma$ being the dissipative strength, 
$\xi$ the strength of the random force, 
$\bm{v}_{ij}$ the relative velocity between particles, and $\theta_{ij}$ a random 
variable from a Gaussian distribution with unit variance.
The fluctuation--dissipation theorem requires that
$\gamma$ and $\xi$ be related by the following relation
\begin{eqnarray}
	\gamma = \frac{\xi ^2}{2 k_B T}.
	\label{eq7}
\end{eqnarray}
The temperature of the system in the simulations was $T=1$ (MDPD units)
while the weight functions for the dissipative and random forces are
\begin{eqnarray}
	\omega^D(r_{ij}) = \left[\omega^R(r_{ij})\right]^2 = 
	\begin{cases}
		\left( 1 - \frac{r_{ij}}{r_c}\right)^2,  & r_{ij} \leq r_{c} \\
		0,   & r_{ij} > r_{c}.
	\end{cases}
	\label{eq8}
\end{eqnarray}

The equations of motion (\textbf{Equation~\ref{eq1}}) are integrated by
a modified velocity-Verlet algorithm in LAMMPS \cite{Plimpton1995,Thompson2022}
with a time step $\Delta t = 0.01$.
The non-bonded interactions between beads are tuned by
varying $A_{ij}$, while repulsive interactions will 
remain constant as $B=35$ \cite{warren2013}. 
In the MDPD model, harmonic bond and angle interactions are employed,
which are mathematically expressed as
\begin{equation}
    E_{bond} = k \left( r_{ij} - r_0 \right)^2
\end{equation}
and
\begin{equation}
    E_{angle} = k_A \left( \theta_{ijk} - \theta_0 \right)^2, 
\end{equation}
respectively. The parameters for the bond
interactions are $k = 150$ and $r_0 = 0.5$, while
for the angles $k_A = 5$, while $\theta_0$ depends
on the specific molecule with values indicated in \textbf{Table \ref{Angle_values}}.

\begin{table}[t]
    \centering
    \begin{tabular}{c|c}
    Angle         & $\theta_0$ \\ \hline
    $Q_0-Q_a-N_a$ & I          \\ 
    $Q_a-N_a-N_a$ & II         \\
    $Q_a-N_a-C_1$ & I          \\
    $N_a-N_a-C_1$ & II         \\
    $N_a-C_1-C_3$ & I          \\
    $N_a-C_1-C_1$ & I          \\
    $C_1-C_3-C_1$ & II         \\
    $C_3-C_1-C_1$ & I          \\
    $C_1-C_1-C_1$ & I          \\
    \end{tabular}
    \caption{$\theta_0$ values for equilibrium angles for triads
    of consecutive CG beads: $180^{\circ}$ (I); $120^{\circ}$ (II).}
    \label{Angle_values}
\end{table}

The MDPD model for the DPPC, POPC, and DOPC is illustrated in 
\textbf{Figure \ref{Lipids_models}}. 
Moreover, the interactions between the different beads are 
presented in \textbf{Table~\ref{interaction_matrix}} \cite{Carnevale2024_MDPD_MARTINI}.
Also, an additional bead-type C3 with respect to the DPPC lipid model
has been considered as in the MD-MARTINI
model in the case of the DOPC and POPC lipids.
Finally, forming the bilayers by using the MDPD-MARTINI model follows the
same simulation protocol
as in the MD-MARTINI simulations with
temperature increase from 300~K to 450~K. 
Typical snapshots for each bilayer case
obtained from MDPD simulations
are presented in \textbf{Figure \ref{MD_MDPD_Bilayers}}
alongside the corresponding MD-MARTINI snapshots.


\begin{table}[t]
    \centering
    \begin{tabular}{c|cccccc}
    & P4  & Q0 & Qa   & Na    & $C_1$ & $C_3$ \\
    \hline
    P4   & I  & I & I & II  & VI & V \\ 
    Q0   & I  & IV & IV & II  & VI & V\\ 
    Qa   & I  & IV & IV & II  & VI & V\\ 
    Na   & II & II & II & III & VI & V \\ 
    $C_1$ & VI  & VI & VI & VI & VI & VI \\ 
    $C_3$ & V & V & V & V & VI & VI \\
    \end{tabular}
    \caption{Interaction matrix for parametrized MDPD beads organized in six interaction levels (I-VI) with corresponding attractive parameters (A):
$-50$ (I); $-43$ (II); $-34$ (III); $-30$ (IV); $-28$ (V); $-26$ (VI). }  
    \label{interaction_matrix} 
\end{table}

\section{Results and Discussion}\label{results}
   
\begin{figure*}[t]
    \centering
    \includegraphics[width=\linewidth]{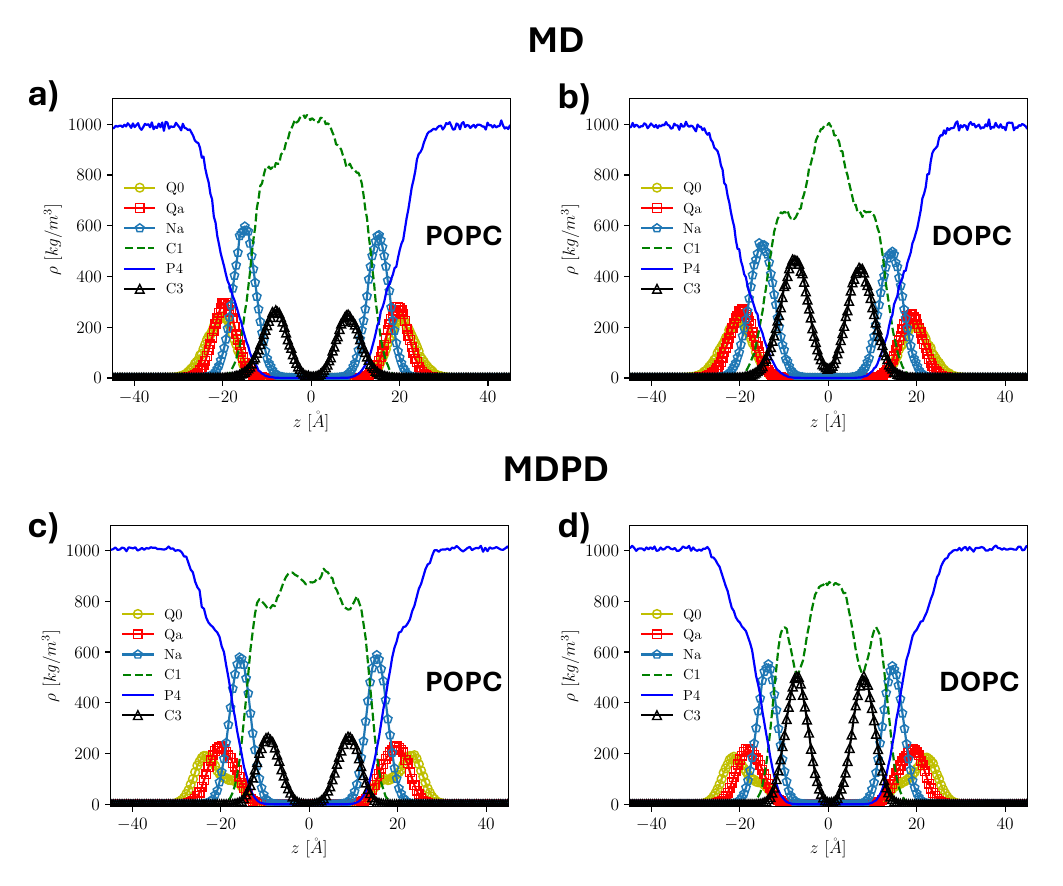}
    \caption{Comparison of bilayer thickness in MD (a, b) and MDPD (c, d) models.
    Results for POPC (a, c) and DOPC (b, d) bilayers consisting of
    512 lipid molecules are shown. The values of bilayer thickness are reported in
    Table~\ref{tab:apl}. }
    \label{density_profiles}
\end{figure*}

\addtolength{\tabcolsep}{6pt} 
\begin{table}[b]
\centering
\caption{MDPD and MD measurements of area per lipid and bilayer thickness for the
different lipid molecules considered. Both methods produce membranes of comparable thickness, while the current MDPD force-field underestimates the APL. }
\begin{tabular}{llll}
  &  & MDPD  & MD   \\ 
  \midrule 
  \multirow{2}{1cm}{DPPC} &  APL        &  $46.2$ \AA$^2$ & $51.6$ \AA$^2$  \\ 
                          &  Thickness  &  $44.2$ \AA     & $44.0$ \AA      \\
                          \midrule 
  \multirow{2}{1cm}{POPC} &  APL        &  $54.3$ \AA$^2$ & $61.6$ \AA$^2$  \\ 
                          &  Thickness  &  $40.1$ \AA     & $39.5$ \AA      \\ 
                          \midrule 
  \multirow{2}{1cm}{DOPC} &  APL        &  $59.1$ \AA$^2$ & $64.7$ \AA$^2$  \\ 
                          &  Thickness  &  $37.8$ \AA     & $38.4$ \AA      \\ 
\end{tabular}
\label{tab:apl} 
\end{table}
\addtolength{\tabcolsep}{-6pt} 



The MD and MDPD MARTINI systems typically consisted of 512 lipids, 
while in certain cases, for example, for determining the spectrum intensities 
of the undulation modes on the bilayer, 
larger systems have been used
(8192 lipids in total). \blu{Moreover, different sizes were tested in our previous work on DPPC lipids and all cases were able to form a stable bilayer \cite{Carnevale2024_MDPD_MARTINI}.}
Firstly, the area per lipid (APL) was examined. This can be determined
by the area a lipid covers in such bilayers, in other words, this is the
area of the bilayer divided by the average number of lipids in a single 
monolayer. In our study, the area of 
a plane aligned tangentially to the bilayer surface in the simulation box was 
divided by half the total number of lipid molecules in order to calculate the
APL. This property depends on 
various factors, such as the lipid type and temperature. In this case,
the influence of the lipid type was examined and a comparison between
the MD and MDPD models was performed. The exact values of APL
for the lipid bilayers considered are reported in Table~\ref{tab:apl}. In particular, we can observe
that MDPD exhibits lower APL values than those from MD simulations. Moreover,
both simulation methods present an increase in APL when going from DPPC--POPC--DOPC. This is expected partly due to the 
different angle involving the C3-type beads (Table~\ref{Angle_values}).

Another fundamental property that has been estimated is the thickness bilayer
for the different lipids for both the MDPD- and MD-MARTINI models. This parameter is 
essential for describing structural aspects of the bilayers and was determined
by identifying the peaks in the density distribution profile of the
choline head-group bead (Q$_0$) as shown in Figure~\ref{density_profiles}.
It is found that both the MD and MDPD models are in agreement with each other
and moreover with experimental data \cite{drabik2020} for all lipid bilayer cases.
The exact values are reported in Table~\ref{tab:apl}. In particular,
it has been found that the thickness decreases as the APL increases \blu{for}
the different systems. 
The structural characteristics expressed through
the density profiles of each bilayer for the
various bead types are similar for both the 
MD and the MDPD models, as shown 
in Figure~\ref{density_profiles}. 

\begin{figure*}[t]
    \centering
    \includegraphics[width=\linewidth]{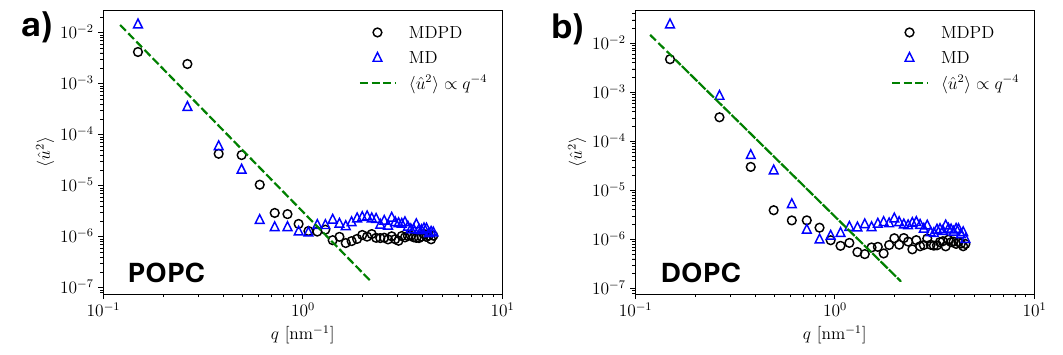}
    \caption{Spectrum intensitites of the undulation modes on the bilayer.
    Long wavelengths obey the $q^{-4}$ behavior and transition around an
    1~nm wavelength to a constant value in both the MD and the MDPD simulations
    is observed for POPC (a) and DOPC (bilayers). Data are obtained for a 
    system of 8192 lipids.}
    \label{bending}
\end{figure*}

\begin{figure*}[t]
    \centering
    \includegraphics[width=\linewidth]{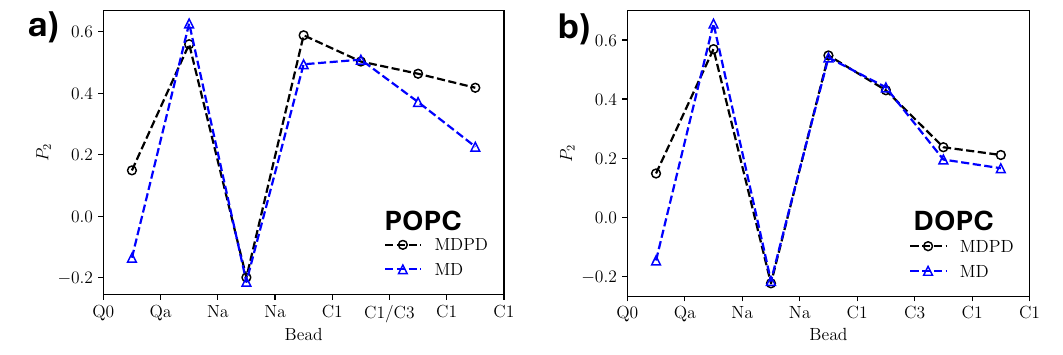}
    \caption{Second-rank order-parameter, $P_2$, calculated for every bond along
    the lipid molecule in the case of POPC and DOPC lipids. Each bilayer
    consisted of 8192 lipids.}
    \label{p2_order_parameter}
\end{figure*}

Lipid bilayers possess elastic properties, which can be expressed by the bending modulus, $k_c$. In particular, this reflects the intensity of undulations
within the lipid membrane and can be obtained by the fluctuation 
spectrum \cite{tarazona2013}. By replicating the bilayers in
the bilayer plane and running short simulations to reach equilibrium
structures, the undulation spectrum was determined. 
Specifically, undulations can be described through a surface
function, $u(x, y)$, which represents the average position of the
two monolayers based on the positions of the phosphate head beads.
Then, the undulation spectrum, $\hat{u}$, can be obtained 
by performing a Fourier transform on the surface function, that is:
\begin{equation}
    \hat{u}(\mathbf{q}) = \frac{1}{N}\int u(\mathbf{x})e^{-i\mathbf{q}\cdot \mathbf{x}} \mathbf{dx}.
    \label{eq:fourier}
\end{equation} 
Moreover, the undulation intensities in
the Fourier space are expected to follow the relation \cite{tarazona2013}
\begin{equation}
    \langle \hat{u}^2 \rangle = \frac{k_B T}{A} \left( 
    \frac{1}{k_cq^4} + \frac{1}{\gamma q^2} \right),
    \label{eq:bendingModulus}
\end{equation}
where $A = L_xL_y$ is the lipid-membrane area, 
while $q = |\mathbf{q}| = \sqrt{q_x^2 + q_y^2}$  is the wavevector length
in the radial direction and $\gamma$ the surface tension. 
Our data, shown in Figure~\ref{bending}, for the DOPC and POPC bilayers exhibit the $q^{-4}$ scaling for
the long wavelength modes, with a very good match between the MD
and the MDPD results. Moreover, in comparison with previous results
for DPPC bilayers \cite{Carnevale2024_MDPD_MARTINI}, here, the $q^{-2}$
dependence is not present due to the application of the zero surface-tension condition. The $q^{-4}$ dashed line was plotted by considering 
the experimentally obtained bending modulus $k_c=5\time 10^{-20}$J \citep{rawicz2000}.

To further compare the MD- and MDPD-MARTINI results, the second-rank order
parameter $P_2$ was computed \cite{Chremos2016}, given by the following relation:
\begin{equation}
    P_2 = \frac{1}{2} \left( 3\langle \cos^2\theta \rangle -1 \right)
    \label{eq:orderParameter}
\end{equation}
Here, $\theta$ is the angle between each bond 
and the bilayer normal direction. Thus, it describes the alignment of the lipid
molecules. In particular, values closer to $P_2=1$ indicate a higher
degree of alignment with the normal direction, 
whereas values around $P_2=0$ rather point to a random orientation.
Finally, values towards $P_2=-0.5$ indicate anti-alignment. 
The obtained results indicate that the MDPD-MARTINI 
model can reproduce the alignment characteristics of
the lipid molecules as in the MD simulations.

\section{Conclusions}
\label{conclusions}

The MDPD-MARTINI-FF was employed for investigating 
structural properties of lipid bilayers with results
being compared to simulations based on the widely 
used MD-MARTINI CG FF. The two models are in very good
agreement in a range of different properties 
discussed in this study, thus providing further 
support in favor of using MDPD models,
since simulations are much faster 
in terms of computational and physical times \cite{Carnevale2024_MDPD_MARTINI}.
This particular feature opens the door for reaching
larger systems and longer times scales in the
simulation of complex systems, covering the full-range of
chemical groups currently available in the MD-MARTINI models.
This will also have the potential of accelerating the acquisition
of physical-based data, which can be used in machine learning or
other artificial-intelligence methods.
\blu{In various applications, the presence of
soft interactions in MDPD models may pose certain
limitations when hard-core interactions might
be necessary. While this remains to be seen, various solutions can be envisioned in such
a scenario beyond the possible use of hybrid models based on MDPD and LJ interactions. In addition, further work with MDPD
models is expected to provide more information regarding
any limitations of the MDPD models for specific systems,
as MDPD models gain in popularity.

In this study,}
the FF parameters of the
C3 MARTINI bead-type has been provided to the MDPD-MARTINI interaction 
matrix \cite{Carnevale2024_MDPD_MARTINI}. 
In view of the transferability of the potential
based on the hydrophobic/hydrophilic characteristics
of chemical groups represented by the CG beads, 
which is an important concept
in the MARTINI coarse-graining approach and demonstrated here in the case
of DPPC, POPC, and DOPC lipids, a range of different systems
can be simulated by the provided set of interactions.
Thus, by following the MARTINI mapping of CG beads to
chemical units additional systems can be investigated for further validations. \blu{Finally, methods, such as the G\={o}MARTINI for proteins\cite{Poma2017}, will
be integrated into the general-purpose MDPD-MARTINI-FF as things progress.}


\section*{Acknowledgments}
This research has been supported by the National Science Centre, Poland, under grant\\ No.\ 2019/34/E/ST3/00232. We gratefully acknowledge Polish high-performance computing infrastructure PLGrid (HPC Center: ACK Cyfronet AGH) for providing computer facilities and support within computational grant no. PLG/2023/016608.

\section*{Data Availability Statement}
Data sets generated during the current study are available from the corresponding author on reasonable request.

\bibliographystyle{spphys}       
\bibliography{bib}   

\begin{thebibliography}{10}
\providecommand{\url}[1]{{#1}}
\providecommand{\urlprefix}{URL }
\expandafter\ifx\csname urlstyle\endcsname\relax
  \providecommand{\doi}[1]{DOI \discretionary{}{}{}#1}\else
  \providecommand{\doi}{DOI \discretionary{}{}{}\begingroup \urlstyle{rm}\Url}\fi

\bibitem{Rapaport_2004}
D.C. Rapaport, \emph{The Art of Molecular Dynamics Simulation}, 2nd edn. (Cambridge University Press, 2004)

\bibitem{Noid2013}
W.G. Noid, J. Chem. Phys. \textbf{139}(9), 090901 (2013).
\newblock \doi{10.1063/1.4818908}.
\newblock \urlprefix\url{https://doi.org/10.1063/1.4818908}

\bibitem{Kmiecik2016}
S.~Kmiecik, D.~Gront, M.~Kolinski, L.~Wieteska, A.E. Dawid, A.~Kolinski, Chem. Rev. \textbf{116}(14), 7898 (2016).
\newblock \doi{10.1021/acs.chemrev.6b00163}

\bibitem{Jin2022}
J.~Jin, A.J. Pak, A.E.P. Durumeric, T.D. Loose, G.A. Voth, J. Chem. Theory Comput. \textbf{18}(10), 5759 (2022).
\newblock \doi{10.1021/acs.jctc.2c00643}

\bibitem{Noid2023}
W.G. Noid, J. Phys. Chem. B \textbf{127}(19), 4174 (2023).
\newblock \doi{10.1021/acs.jpcb.2c08731}

\bibitem{Marrink2004}
S.J. Marrink, A.H. de~Vries, A.E. Mark, J. Phys. Chem. B \textbf{108}(2), 750 (2004).
\newblock \doi{10.1021/jp036508g}

\bibitem{Marrink2007}
S.J. Marrink, H.J. Risselada, S.~Yefimov, D.P. Tieleman, A.H. de~Vries, J. Phys. Chem. B \textbf{111}(27), 7812 (2007).
\newblock \doi{10.1021/jp071097f}

\bibitem{Marrink2013}
S.J. Marrink, D.P. Tieleman, Chem. Soc. Rev. \textbf{42}, 6801 (2013).
\newblock \doi{10.1039/C3CS60093A}.
\newblock \urlprefix\url{http://dx.doi.org/10.1039/C3CS60093A}

\bibitem{Alessandri2021}
R.~Alessandri, F.~Gr\"unewald, S.J. Marrink, Adv. Mat. \textbf{33}(24), 2008635 (2021).
\newblock \doi{https://doi.org/10.1002/adma.202008635}.
\newblock \urlprefix\url{https://onlinelibrary.wiley.com/doi/abs/10.1002/adma.202008635}

\bibitem{Monticelli2008}
L.~Monticelli, S.K. Kandasamy, X.~Periole, R.G. Larson, D.P. Tieleman, S.J. Marrink, J. Chem. Theory Comput. \textbf{4}(5), 819 (2008).
\newblock \doi{10.1021/ct700324x}

\bibitem{Periole2009}
X.~Periole, M.~Cavalli, S.J. Marrink, M.A. Ceruso, J. Chem. Theory Comput. \textbf{5}(9), 2531 (2009).
\newblock \doi{10.1021/ct9002114}

\bibitem{deJong2013}
D.H. de~Jong, G.~Singh, W.F.D. Bennett, C.~Arnarez, T.A. Wassenaar, L.V. Sch\"afer, X.~Periole, D.P. Tieleman, S.J. Marrink, J. Chem. Theory Comput. \textbf{9}(1), 687 (2013).
\newblock \doi{10.1021/ct300646g}

\bibitem{Poma2017}
A.B. Poma, M.~Cieplak, P.E. Theodorakis, J. Chem. Theory Comput. \textbf{13}(3), 1366 (2017).
\newblock \doi{10.1021/acs.jctc.6b00986}

\bibitem{Lee2009}
H.~Lee, A.H. de~Vries, S.J. Marrink, R.W. Pastor, J. Phys. Chem. B \textbf{113}(40), 13186 (2009).
\newblock \doi{10.1021/jp9058966}

\bibitem{Lopez2009}
C.A. L\'opez, A.J. Rzepiela, A.H. de~Vries, L.~Dijkhuizen, P.H. Hünenberger, S.J. Marrink, J. Chem. Theory Comput. \textbf{5}(12), 3195 (2009).
\newblock \doi{10.1021/ct900313w}

\bibitem{Lopez2013}
C.A. L\'opez, Z.~Sovova, F.J. van Eerden, A.H. de~Vries, S.J. Marrink, J. Chem. Theory Comput. \textbf{9}(3), 1694 (2013).
\newblock \doi{10.1021/ct3009655}

\bibitem{Chakraborty2021}
S.~Chakraborty, K.~Wagh, S.~Gnanakaran, C.A. López, Glycobiology \textbf{31}(7), 787 (2021).
\newblock \doi{10.1093/glycob/cwab017}.
\newblock \urlprefix\url{https://doi.org/10.1093/glycob/cwab017}

\bibitem{Uusitalo2015}
J.J. Uusitalo, H.I. Ingólfsson, P.~Akhshi, D.P. Tieleman, S.J. Marrink, J. Chem. Theory Comput. \textbf{11}(8), 3932 (2015).
\newblock \doi{10.1021/acs.jctc.5b00286}

\bibitem{Uusitalo2017}
J.J. Uusitalo, H.I. Ingólfsson, S.J. Marrink, I.~Faustino, Biophys. J. \textbf{113}(2), 246 (2017).
\newblock \doi{https://doi.org/10.1016/j.bpj.2017.05.043}

\bibitem{Yesylevskyy2010}
S.O. Yesylevskyy, L.V. Schäfer, D.~Sengupta, S.J. Marrink, PLOS Comput. Biol. \textbf{6}(6), 1 (2010).
\newblock \doi{10.1371/journal.pcbi.1000810}.
\newblock \urlprefix\url{https://doi.org/10.1371/journal.pcbi.1000810}

\bibitem{Vainikka2021}
P.~Vainikka, S.~Thallmair, P.C.T. Souza, S.J. Marrink, ACS Sustain. Chem. Eng. \textbf{9}(51), 17338 (2021).
\newblock \doi{10.1021/acssuschemeng.1c06521}

\bibitem{Grunewald2020}
F.~Gr\"unewald, P.C.T. Souza, H.~Abdizadeh, J.~Barnoud, A.H. de~Vries, S.J. Marrink, J. Chem. Phys. \textbf{153}(2), 024118 (2020).
\newblock \doi{10.1063/5.0014258}.
\newblock \urlprefix\url{https://doi.org/10.1063/5.0014258}

\bibitem{Sami2023}
S.~Sami, S.J. Marrink, J. Chem. Theory Comput. \textbf{19}(13), 4040 (2023).
\newblock \doi{10.1021/acs.jctc.2c01186}

\bibitem{Souza2021}
P.C.T. Souza, R.~Alessandri, J.~Barnoud, S.~Thallmair, I.~Faustino, F.~Gr\"unewald, I.~Patmanidis, H.~Abdizadeh, B.M.H. Bruininks, T.A. Wassenaar, P.C. Kroon, J.~Melcr, V.~Nieto, V.~Corradi, H.M. Khan, J.~Doma\'nski, M.~Javanainen, H.~Martinez-Seara, N.~Reuter, R.B. Best, I.~Vattulainen, L.~Monticelli, X.~Periole, D.P. Tieleman, A.H. de~Vries, S.J. Marrink, Nat. Methods \textbf{18}, 382 (2021).
\newblock \doi{10.1038/s41592-021-01098-3}

\bibitem{Berendsen1995}
H.~Berendsen, D.~{van der Spoel}, R.~{van Drunen}, Comput. Phys. Commun. \textbf{91}(1), 43 (1995).
\newblock \doi{https://doi.org/10.1016/0010-4655(95)00042-E}.
\newblock \urlprefix\url{https://www.sciencedirect.com/science/article/pii/001046559500042E}

\bibitem{Carnevale2024_MDPD_MARTINI}
L.H. Carnevale, P.E. Theodorakis, Eur. Phys. J. Plus \textbf{139}, 539 (2024).
\newblock \doi{10.1140/epjp/s13360-024-05362-1}

\bibitem{Pagonabarraga2001}
I.~Pagonabarraga, D.~Frenkel, J. Chem. Phys. \textbf{115}(11), 5015 (2001).
\newblock \doi{10.1063/1.1396848}

\bibitem{warren2003}
P.B. Warren, Phys. Rev. E \textbf{68}, 066702 (2003).
\newblock \doi{10.1103/PhysRevE.68.066702}

\bibitem{zhao2017}
J.~Zhao, S.~Chen, N.~Phan-Thien, Mol. Sim. \textbf{44}(3), 213 (2017).
\newblock \doi{10.1080/08927022.2017.1364379}

\bibitem{zhao2021-review}
J.~Zhao, S.~Chen, K.~Zhang, Y.~Liu, Phys. Fluids \textbf{33}(11), 112002 (2021).
\newblock \doi{10.1063/5.0065538}

\bibitem{zhao2021}
C.~Zhao, J.~Zhao, T.~Si, S.~Chen, Phys. Fluids \textbf{33}(11), 112004 (2021).
\newblock \doi{10.1063/5.0066982}

\bibitem{Han2021}
Y.~Han, J.~Jin, G.A. Voth, J. Chem. Phys. \textbf{154}(8), 084122 (2021).
\newblock \doi{10.1063/5.0035184}

\bibitem{Zhu2021}
Y.~Zhu, X.~Bai, G.~Hu, Biophys. J. \textbf{120}(21), 4751 (2021).
\newblock \doi{https://doi.org/10.1016/j.bpj.2021.09.031}.
\newblock \urlprefix\url{https://www.sciencedirect.com/science/article/pii/S0006349521007931}

\bibitem{Ghoufi2011}
A.~Ghoufi, P.~Malfreyt, Phys. Rev. E \textbf{83}, 051601 (2011).
\newblock \doi{10.1103/PhysRevE.83.051601}

\bibitem{lipids_tutorial}
 \urlprefix\url{www.cgmartini.nl/index.php/tutorial/37-tutorial2/356-tutorial-lipids-new-lipids}.
\newblock \url{www.cgmartini.nl/index.php/tutorial/37-tutorial2/356-tutorial-lipids-new-lipids}, accessed 25.09.2023

\bibitem{Plimpton1995}
S.~Plimpton, J. Comp. Phys. \textbf{117}, 1 (1995)

\bibitem{Thompson2022}
A.P. Thompson, H.M. Aktulga, R.~Berger, D.S. Bolintineanu, W.M. Brown, P.S. Crozier, P.J. {in 't Veld}, A.~Kohlmeyer, S.G. Moore, T.D. Nguyen, R.~Shan, M.J. Stevens, J.~Tranchida, C.~Trott, S.J. Plimpton, Comput. Phys. Commun. \textbf{271}, 108171 (2022).
\newblock \doi{https://doi.org/10.1016/j.cpc.2021.108171}.
\newblock \urlprefix\url{https://www.sciencedirect.com/science/article/pii/S0010465521002836}

\bibitem{lipids_tutorial_moltemplate}
 \urlprefix\url{https://www.moltemplate.org/visual\_examples.html}.
\newblock \url{https://www.moltemplate.org/visual\_examples.html}, accessed 27.09.2023

\bibitem{Espanol1995}
P.~Español, P.~Warren, Europhys. Lett. (EPL) \textbf{30}(4), 191 (1995).
\newblock \doi{10.1209/0295-5075/30/4/001}.
\newblock \urlprefix\url{https://dx.doi.org/10.1209/0295-5075/30/4/001}

\bibitem{Vanya2018}
P.~Vanya, P.~Crout, J.~Sharman, J.A. Elliott, Phys. Rev. E \textbf{98}, 033310 (2018).
\newblock \doi{10.1103/PhysRevE.98.033310}

\bibitem{Carnevale2023}
L.H. Carnevale, P.~Deuar, Z.~Che, P.E. Theodorakis, Phys. Fluids \textbf{35}(7), 074108 (2023).
\newblock \doi{10.1063/5.0157752}.
\newblock \urlprefix\url{https://doi.org/10.1063/5.0157752}

\bibitem{Ng2025}
K.L. Ng, L.H. Carnevale, M.~Klamka, P.~Deuar, T.~Bobinski, P.E. Theodorakis, Physics of Fluids \textbf{37}(1), 014121 (2025).
\newblock \doi{10.1063/5.0250009}

\bibitem{Hendrikse2023}
R.L. Hendrikse, C.~Amador, M.R. Wilson, Soft Matter \textbf{19}, 3590 (2023).
\newblock \doi{10.1039/D3SM00276D}.
\newblock \urlprefix\url{http://dx.doi.org/10.1039/D3SM00276D}

\bibitem{Carnevale2024}
L.H. Carnevale, P.~Deuar, Z.~Che, P.E. Theodorakis, Phys. Fluids \textbf{36}(3), 033301 (2024).
\newblock \doi{10.1063/5.0198154}.
\newblock \urlprefix\url{https://doi.org/10.1063/5.0198154}

\bibitem{Theodorakis2011}
P.E. Theodorakis, W.~Paul, K.~Binder, Eur. Phys. J. E \textbf{34}, 52 (2011).
\newblock \doi{10.1140/epje/i2011-11052-5}

\bibitem{Espanol2017}
P.~Español, P.B. Warren, J. Chem. Phys. \textbf{146}(15), 150901 (2017).
\newblock \doi{10.1063/1.4979514}

\bibitem{Yoshimoto2013}
Y.~Yoshimoto, I.~Kinefuchi, T.~Mima, A.~Fukushima, T.~Tokumasu, S.~Takagi, Phys. Rev. E \textbf{88}, 043305 (2013).
\newblock \doi{10.1103/PhysRevE.88.043305}.
\newblock \urlprefix\url{https://link.aps.org/doi/10.1103/PhysRevE.88.043305}

\bibitem{Li2014}
Z.~Li, X.~Bian, B.~Caswell, G.E. Karniadakis, Soft Matter \textbf{10}, 8659 (2014).
\newblock \doi{10.1039/C4SM01387E}

\bibitem{Lavagnini2021}
E.~Lavagnini, J.L. Cook, P.B. Warren, C.A. Hunter, J. Phys. Chem. B \textbf{125}(15), 3942 (2021).
\newblock \doi{10.1021/acs.jpcb.1c00480}

\bibitem{Trofimov2005}
S.Y. Trofimov, E.L.F. Nies, M.A.J. Michels, J. Chem. Phys. \textbf{123}(14), 144102 (2005).
\newblock \doi{10.1063/1.2052667}

\bibitem{warren2013}
P.B. Warren, Phys. Rev. E \textbf{87}, 045303 (2013).
\newblock \doi{10.1103/PhysRevE.87.045303}

\bibitem{drabik2020}
D.~Drabik, G.~Chodaczek, S.~Kraszewski, M.~Langner, Langmuir \textbf{36}(14), 3826 (2020).
\newblock \doi{10.1021/acs.langmuir.0c00475}.
\newblock \urlprefix\url{https://doi.org/10.1021/acs.langmuir.0c00475}

\bibitem{tarazona2013}
P.~Tarazona, E.~Chacón, F.~Bresme, J. Chem. Phys. \textbf{139}(9), 094902 (2013).
\newblock \doi{10.1063/1.4818421}

\bibitem{rawicz2000}
W.~Rawicz, K.~Olbrich, T.~McIntosh, D.~Needham, E.~Evans, Biophys. J. \textbf{79}(1), 328 (2000).
\newblock \doi{https://doi.org/10.1016/S0006-3495(00)76295-3}.
\newblock \urlprefix\url{https://www.sciencedirect.com/science/article/pii/S0006349500762953}

\bibitem{Chremos2016}
A.~Chremos, P.E. Theodorakis, Polymer \textbf{97}, 191 (2016).
\newblock \doi{https://doi.org/10.1016/j.polymer.2016.05.034}

\end{thebibliography}

\end{document}